\documentclass[twocolumn,aps,10pt,pra,nofootinbib,dvipsnames,superscriptaddress]{revtex4}
	
\usepackage{graphicx}
\usepackage{bm}
\usepackage{amsmath,hyperref,amsthm,ulem}
\usepackage{amssymb}
\usepackage{appendix}
\usepackage{amsfonts}
\usepackage{fancyhdr}
\usepackage{slashed}
\usepackage{latexsym,epsfig,bbm}
\usepackage{qcircuit}
\usepackage{algorithm}

\usepackage{pdfcomment}

\usepackage[inline]{enumitem}

 \theoremstyle{plain}

\newtheorem{oracle}{Input Assumption}

\newcommand{\bra}[1]{\langle{#1}|}
\newcommand{\ket}[1]{|{#1}\rangle}

\newcommand{\braket}[2]{\langle{#1}|{#2}\rangle}

\newcommand{\Tr}{\mathrm{Tr}}
\usepackage{color}
\definecolor{blue}{rgb}{0,0.2,1}

\definecolor{red}{rgb}{0.9,0,0}

\newcommand{\Ord}[1]{{O}\left( #1 \right)}

\begin{document}
\title{Compiling basic linear algebra subroutines for quantum computers}
%\title{Quantum matrix algebra toolbox}

\author{Liming Zhao}
\affiliation{Singapore University of Technology and Design,  8 Somapah Road, Singapore 487372}
\affiliation{Centre for Quantum Technologies, National University of Singapore, 3 Science Drive 2, Singapore 117543}
\author{Zhikuan Zhao}
\affiliation{Singapore University of Technology and Design,  8 Somapah Road, Singapore 487372}
\affiliation{Centre for Quantum Technologies, National University of Singapore, 3 Science Drive 2, Singapore 117543}
\author{Patrick Rebentrost}
\affiliation{Centre for Quantum Technologies, National University of Singapore, 3 Science Drive 2, Singapore 117543}
\author{Joseph Fitzsimons}
\affiliation{Singapore University of Technology and Design,  8 Somapah Road, Singapore 487372}
\affiliation{Centre for Quantum Technologies, National University of Singapore, 3 Science Drive 2, Singapore 117543}
\affiliation{Horizon Quantum Computing, 79 Ayer Rajah Crescent, BASH \#03-01, Singapore 139955}

\date{\today}

\begin{abstract}

Efficiently processing basic linear algebra subroutines is of great importance for a wide range of computational problems. In this paper, we consider techniques to implement matrix functions on a quantum computer, which are composed of  basic matrix operations on a set of matrices.
These matrix operations include addition, multiplication, Kronecker sum, tensor product,  Hadamard product, and single-matrix functions.  We discuss the composed matrix functions in terms of the estimation of scalar quantities such as inner products,  trace,  determinant and Schatten p-norms. We thus provide a framework for compiling instructions for linear algebraic computations into gate sequences on actual quantum computers. 
\end{abstract}

\maketitle

Quantum algorithms promise exponential speed-ups over their best known classical counterparts for certain problems. An example is the seminal Shor's algorithm that is able to find the prime factors of a given integer efficiently \citep{shor1994algorithms}. Because of such breakthroughs, substantial effort has been devoted to the study of quantum computation over the last several decades. There already exist several quantum processors with a limited number of qubits and ongoing efforts promise to increase the quantity, quality, and connectivity of the qubits. As the impressive development of quantum computers continues, it is highly desirable to study quantum algorithms for a wide range of applications.   

Matrix algebra is widely used in almost every area of science and technology. In many practical problems, the size of the input data, often in the form of vectors or matrices, is a bottleneck for efficient computations. The usual classical algorithms, for example for finding eigenvalues, can be unsuitable for large matrices. Efficient quantum algorithms for linear algebra and matrix inversion have been investigated \cite{itakura2005quantum,buhrman2006quantum,janzing2007simple,harrow2009quantum,lloyd2014quantum}, with a quantum singular value transformation being a recent innovation \cite{Gilyen2018}. These algorithms have found widespread application in quantum machine learning \citep{rebentrost2014quantum,zhao2015quantum,kimmel2017hamiltonian,zhao2018gaussian}. Many problems in machine learning and data processing involve a large number of matrix operations and it is useful to concatenate such operations effectively on a quantum computer. In an ideal scenario, a user of a quantum computer only declares a series of instructions on a set of matrices in a simple language, which is then compiled into a sequence of quantum gates for the quantum computer. An important intermediate step is to provide easy-to-compose techniques for matrix linear algebra.

In this paper, we present a scheme for compiling elementary linear algebraic operations on a quantum computer for a set of complex matrices, given the availability of particular unitaries generated from these matrices. The elementary matrix operations we consider here are matrix addition, multiplication, Kronecker sum, tensor product and Hadamard product, as well as arbitrary real single-matrix functions $h(A)$ that act on the eigenvalues of a matrix $A$. We refer to this scheme as the Quantum Matrix Algebra Toolbox (QMAT) in the following. Specifically, given a set of matrices $\{A_j\}$, we embed these matrices into a set of Hermitian matrices and assume as input a given set of unitary operators generated by the embedding matrices. In addition, we are given a matrix function  $f(\{A_j\})$ that can be divided into tiers of sub-functions, where each sub-function is one of the QMAT operations.  We are then able to compose the matrix operations and compile the tiers of sub-functions into a sequence of operations for the quantum computer. This results in an approximation of a unitary operator which encodes the computation of the matrix function $f(\{A_j\})$ in a composable manner. Such a toolbox is of limited use if the desired outcome is a large vector or matrix since in this case obtaining a complete classical representation is often resource intensive \cite{carmeli2016stable,goyeneche2015five}. Hence we discuss scalar outputs, such as the trace, determinant, and Schatten $p$-norm of $f(\{A_j\})$, which often can be obtained efficiently. We  note recent efficient development of logarithmic time classical algorithms for various machine learning tasks \cite{Tang2018, Tang2018_2,Gilyen2018samp, Chia2018} based on preprocessed classical data. In contrast, the scheme presented here assumes only the availability of particular unitaries and is inherently BQP complete, as it allows for the encoding of arbitrary quantum computations.

\section{Embedding}

We assume a set of matrices $\{A_j\}$, $j=1,\dots,J$, with $A_j\in \mathbb C^{N\times N}$. 
The matrices are not required to be Hermitian.
The restriction to square matrices is for consistency in the output of the basic matrix operations. Non-square matrices can always be padded by zeros to turn them into square matrices. The results of the QMAT operations for the original non-square matrices can be obtained by removing the corresponding zeros from the results of the extended matrices.
 In the QMAT setting, the non-Hermitian matrices $\{A_j\}$ are embedded into Hermitian matrices. 
 A Hermitian matrix can be interpreted as a Hamiltonian operator in quantum mechanics and Hamiltonian simulation techniques can be used to perform matrix operations.  
 Given a matrix $A\in \mathbb{C}^{N\times N}$,  we define the following embedding matrices $X_1(A) $, $X_2(A)$, and $X_3(A)$  as
 \begin{align}\label{eqEmbeddingMatrices}
&X_1(A)=R_1\otimes A+R_1^\dagger\otimes A^\dagger,	\nonumber\\	
&X_2(A)=R_2\otimes A+R_2^\dagger\otimes A^\dagger,	\nonumber\\
&X_3(A)=R_3\otimes A+R_3^\dagger\otimes A^\dagger,
\end{align}
where the $R_1$, $R_2$ and $R_3$ are $3\times3$ matrices, given by
\begin{align}
R_1&=\left[\begin{array}{ccc}
0 & 1 & 0\\
0 & 0 & 0\\
0 & 0 & 0
\end{array} \right],~
R_2=\left[\begin{array}{ccc}
0 & 0 & 0\\
0 & 0 & 1\\
0 & 0 & 0
\end{array} \right],~  
R_3=\left[\begin{array}{ccc}
0 & 0 & 1\\
0 & 0 & 0\\
0 & 0 & 0
\end{array} \right].\nonumber \\
\end{align}
Analogously, we embed the vectors  $\bold{x}\in \mathbb{C}^{N} , \bold{y} \in \mathbb{C}^{N}$
 into the vectors $V_1(\bold{{x}})$, $V_2(\bold{{y}})$, which are defined as 
\begin{align}\label{eqEmbeddingMatrices}
&V_1(\bold{{x}})=\bold{r}_1 \otimes \bold{x}, \:
V_2(\bold{{y}})=\bold{r}_2\otimes\bold{y},
\end{align}
where  $\bold{r}_1$ and $\bold{r}_2$ are $3$-dimensional vectors
\begin{align}
\bold{r}_1&=\left[\begin{array}{ccc}
 1&
 0&
 0
\end{array} \right]^T,~
\bold{r}_2=\left[\begin{array}{ccc}
 0&
 0&
 1
\end{array} \right]^T.\nonumber
\end{align}
It follows that
\begin{eqnarray}
V_1(\bold{{x}})^\dagger X_3(A) V_2(\bold{{y}})=\bold{x}^\dagger A\bold{y}. 
\end{eqnarray}
Hence the value of $ \bold{x}^\dagger A\bold{y}$ can be obtained by calculating the inner product of the corresponding embedded matrix and vectors. 

We now define the following three permutation operators, $P_1$, $P_2$ and $P_3$ 
\begin{eqnarray}
P_1=\left[\begin{array}{ccc}
0 & 0 & I\\
I & 0 & 0\\
0 & I & 0
\end{array} \right],~
P_2=\left[\begin{array}{ccc}
0 & I & 0\\
I & 0 & 0\\
0 & 0 & I
\end{array} \right],~
P_3=\left[\begin{array}{ccc}
I & 0 & 0\\
0 & 0 & I\\
0 & I & 0
\end{array} \right], \nonumber
\end{eqnarray}
where $I$ denotes the $N \times N$ identity matrix and $0$ denotes the $N \times N$ matrix of zeros. Then the embedding matrices $X_1(A)$, $X_2(A)$ and $X_3(A)$ can be transformed into one another by applying these permutation operators
\begin{eqnarray}
P_i X_i(A)P_i^{\dagger}&=&X_{(i+1)\text{mod}\, 3}(A), ~~(i=1,2,3),\label{permutation}
\end{eqnarray}
which also holds for the exponentiated $X_i(A)$ since $P_1$, $P_2$ and $P_3$ are unitary.
The transformations above can also be achieved in the reverse direction.

\section{Quantum algorithms for matrix operations}

We now discuss the quantum algorithms for matrix operations. We start with the required data input model and continue with obtaining the desired output.
We then show how one can simulate the unitary operator generated by $X_3(f(\{A_j\}))$ when given access to a set of input unitary operators generated by $X_3(A_j)$. We will describe the QMAT quantum subroutines for the exponentiation of matrix addition, multiplication, Kronecker sum, tensor product and Hadamard product of matrices $A_1,A_2\in \mathbb{C}^{N\times N}$, as well as single matrix functions.  

\subsection{Data input}
We first summarize the data input required in this work. 
\begin{oracle}\label{oracleUnitaries}
Given matrices $A_j$ for $j=1,\dots,J$, let $ \Vert X_3(A_j) \Vert_{\max} \tau = O(1)$, where $\tau$ is a  time parameter\footnote{We assume that $\tau$ can be made arbitrarily small.}. Assume access to the unitaries  $e^{iX_3(A_j)\tau}$. In addition, given arbitrary ancilla qubits assume access to the controlled unitaries $e^{i \ket 1 \bra 1 \otimes X_3(A_j)\tau}$.
\end{oracle}
Here, we use the maximum element norm $\Vert A\Vert_{\max}=\max_{ij} \vert A_{ij} \vert$.
If the matrices $A_j$, and hence $X_3(A_j)$, are sparse and readily stored in a sparse matrix data structure, such unitaries are provided by using well-studied quantum walk techniques \cite{berry2007efficient,Berry2012}. These techniques have been continuously improved to a nearly optimal query complexity of $\Ord{\gamma \frac{\log(\gamma/\epsilon)}{\log\log(\gamma/\epsilon)}}$ \cite{berry2015hamiltonian}, employing the technique of linear combination of unitaries \cite{berry2015truncated}. Here, $\gamma=d \tau \Vert H\Vert_{\max}$, where  $d$ is the sparsity and $\epsilon$ is the desired accuracy in operator norm.   The gate complexity of this approach scales as $\Ord{\log N + \log^{5/2}(d/\epsilon)}$ for each query. An optimal query complexity of  $\Ord{\gamma+ \frac{\log(1/\epsilon)}{\log\log(1/\epsilon)}}$  has been achieved by using quantum signal processing \cite{low2017optimal}. 
Additionally, if the matrix is provided as quantum density matrix we can use the quantum state exponentiation technique \cite{lloyd2014quantum}.

If we are interested in inner products such as $\bold{x}^\dagger f(\{ A_j\})\bold{y}$, we require the preparation of quantum states corresponding to $\bold{x}$ and $\bold{y}$.  This input assumption is not required if we are interested in the inherent scalar quantities of the output matrices such as the trace or determinant.
\begin{oracle}\label{oracleStates}
Assume routines that prepare the quantum states $\ket{{V_i(\bold{v})}}=V_i(\bold{v})/{\vert \bold{v} \vert}$ for the classical vectors $\bold{v} = \bold{x},\bold{y}$ and $i=1,2$.
\end{oracle}
Such data access can be provided if the data are stored in a quantum random access memory (QRAM) as discussed in \cite{giovannetti2008quantum,giovannetti2008architectures}.

\subsection{Output}
\label{secOutput}

We start with a description of a method for calculating an inner product $\bra{{V_1(x)}}X_3(f(\{A_j\}))\ket{{V_2(y)}}$.  
A method for estimating inner products of the form $\braket{V_1(x)}{V_2(y)}$ was presented in \cite{zhao2015quantum}, which we describe in Appendix \ref{appendixQIP}. This procedure can be extended to compute $\bra{V_1(x)}X_3(f(\{A_j\}))\ket{V_2(y)}$ by performing the matrix-vector multiplication $X_3(f(\{A_j\}))\ket{V_2(y)}$, 
analogous to the algorithms presented in \cite{harrow2009quantum,Wiebe2012}. If given access to the controlled unitary $e^{i \ket 1 \bra 1 \otimes X_3(f(\{A_j\}))t}$, we can perform the quantum phase estimation algorithm \cite{kitaev1995quantum} using $\ket{V_2(y)}$ as the input state. This phase estimation results in 
\begin{equation}
\textstyle\sum\nolimits_{k=0}^{N-1}\gamma_k\ket{\mu_k(f)}\ket{\tilde \lambda_k(f)},
\end{equation}
where $\tilde \lambda_k(f)$ approximates the eigenvalues $\lambda_k(f) \equiv \lambda_k(X_3(f(\{A_j\})))$ of the matrix 
$X_3(f(\{A_j\}))$ and $\ket{\mu_k(f)}$ are the associated eigenvectors. We have defined  $\gamma_k=\braket{\mu_k(f)}{V_2(y)}$.
The phase estimation requires a runtime $\Ord{1/\epsilon_\lambda}$ for an $\epsilon_\lambda$ approximation of the eigenvalues \cite{kitaev1995quantum}. 

Subsequently, we perform a controlled rotation of an ancilla register initialized in $\ket 0$ conditioned on the eigenvalue register, which acts on 
$\ket{\lambda_k(f)} \ket 0$ as 
\begin{align}
\ket{\lambda_k(f)}\left(\sqrt{1-c^2 \lambda_k^2(f)}\ket{0}+ {c}\lambda_k(f)\ket{1}\right).\label{HHL1}
\end{align}
The constant $c$ is chosen such that ${c}{\lambda_k(f)} \leq1$ for all $k$. We then uncompute the eigenvalue register and measure the ancilla qubit. A measurement result of $\ket{1}$ leaves the remaining system in the desired output proportional to
\begin{eqnarray}
\textstyle\sum\nolimits_{k=0}^{N-1} \gamma_k \lambda_k(f) \ket{\mu_k(f)}=X_3(f(\{A_j\}))\ket{V_2(y)}.\label{HHL2}
\end{eqnarray}
The success probability of the measurement is given by $\sum_k \gamma^2_k c^2 \lambda_k^2$.

We continue with the QMAT subroutines.
We present the subroutines for convenience using the unitaries without the control register, see Input Assumption
\ref{oracleUnitaries}.  We note that the subroutines can be made controlled by simply using the corresponding controlled unitaries. 

%----------------------------------------------------------------------------------
%% sub-programs
%----------------------------------------------------------------------------------

%----------------------------------------------------------------------------------
%% addition
%----------------------------------------------------------------------------------

\subsection{Addition}

The addition of matrices via Trotter-based methods is well-known and we discuss it here for completeness in the context of the embedding matrices.
Given access to $e^{iX_3(A_1)\tau}$ and $e^{iX_3(A_2)\tau}$,  we are able to approximate $e^{i(X_3(A_1)+X_3(A_2))t}$ with bounded error by using the Trotter formula, see Subroutine \ref{algoadd}. Hence given access to $e^{iX_3(A_1)t/n}$ and $e^{iX_3(A_2)t/n}$, a constant $\epsilon_1$-error approximation of $e^{i(X_3(A_1)+X_3(A_2))t}$ can be achieved in $n=O({t^2/\epsilon_1})$ steps.
The spectral norm $\Vert \cdot \Vert$ is used for quantifying the error.
It is worth mentioning that 
the constant error can be reduced with the higher-order Suzuki-Trotter formula \cite{childs2010relationship,berry2007efficient,childs2004quantum}.
For example, given further access to $e^{iX_3(A_i)t/2n}$, the same error can be achieved in $n=O({t^\frac{3}{2}/\sqrt{\epsilon_1}})$ steps using longer sequences. 
Note that our Input Assumption \ref{oracleUnitaries} allows for arbitrary-$\tau$ unitaries and thus for all higher-order refinements.

\begin{algorithm}[!htbp]
\floatname{algorithm}{Subroutine}
\caption{ Exponentiation of matrix addition} \label{algoadd}
\begin{flushleft}
\textbf{Input:} A set of unitary operators according to Input Assumption \ref{oracleUnitaries} for $A_j$ with $j=1,2$, a desired error parameter $\epsilon_1$, and desired simulation time $t$.\\
\textbf{Output:} Operator 
$\mathcal U_{\rm add}(t)$ which satisfies 
\begin{equation}
\left \Vert \mathcal U_{\rm add}(t)-e^{iX_3(A_1+A_2)t} \right \Vert \le \epsilon_1.
\end{equation} 
\textbf{Procedure:}  Sequentially apply $e^{iX_3(A_1) t/n}$ and $e^{iX_3(A_2) t/n}$  for a total of $n$ consecutive times as follows: 
\begin{equation}
\mathcal U_{\rm add}(t)=  \left( e^{iX_3(A_1)t/n}e^{iX_3(A_2) t/n}\right)^n,
\end{equation}
where the number of applications of the unitaries with $\tau = t/n$ is proportional to $n=O({t^2/\epsilon_1})$.
\end{flushleft}
\end{algorithm}

%----------------------------------------------------------------------------------
%% multiplication
%----------------------------------------------------------------------------------

\subsection{Multiplication}

One way of performing matrix multiplication applied to a quantum state is via two consecutive phase estimations and controlled rotations. Here, we present a version in the QMAT setting by using the commutator Lie formula.
We show how to carry out a matrix multiplication by using the commutator 
\begin{align}
[X_1(A_1), X_2(A_2)] &= \left[\begin{array}{ccc}
0 & 0 & M\\
0 & 0 & 0\\
-M^\dagger & 0 & 0
\end{array} \right],
\end{align}
where $M=A_1A_2$. Note that the result of the commutator shown in the above equation is not a Hermitian matrix, so we  construct a Hermitian matrix via an imaginary factor
\begin{align}
i[X_1(A_1), X_2(A_2) ]=X_3(iM)\label{eqEmbedCommutator}.
\end{align}
It follows that
\begin{align}
e^{iX_3(iM)}=e^{-[X_1(A_1), X_2(A_2) ]}=e^{[iX_1(A_1), iX_2(A_2) ]}. 
\end{align}
However, our goal is to embed the matrix $M$  instead of $iM$. Fortunately, the term  $e^{iX_3(M)}$ can be obtained from $e^{iX_3(iM)}$ by using an unitary operator
$$U_1=\left[\begin{array}{ccc}
\sqrt{-i}I & 0 & 0\\
0 & I & 0\\
0 & 0 & \sqrt{i}I
\end{array} \right],$$
it follows that,
\begin{align}
e^{iX_3(M)}&=U_1e^{iX_3(iM)}U_1^\dagger. \label{equproduct3}
\end{align}

Based on  Eq.~(\ref{equproduct3}), we present a quantum subroutine to find the exponentiation of the product of matrices $A_1$ and $A_2$ in subroutine \ref{algoproduct}.  For simulating the commutator, 
 we require the sequence
\begin{eqnarray}
\tilde{l}(x_1,x_2)&:=& e^{x_1}e^{x_2}e^{-x_1}e^{-x_2} e^{-x_1}e^{-x_2}e^{x_1}e^{x_2} 
\end{eqnarray}
where $x_1$ and $x_2$ are placeholders for arbitrary matrices. 
As we show in Appendix \ref{appendixHS}, the error of $\tilde{l}(X_1(A_1)t/n,X_2(A_2)t/n)^{n^2/2t} $ for simulating the commutator is bounded by  $\epsilon_2=O(t^3/n^2)$  which requires $n'=O(t^2/\epsilon_2)$ applications of $e^{iX_1(A_1)t/n}$ and $e^{iX_2(A_2)t/n}$.  These unitaries are obtained from Input Assumption \ref{oracleUnitaries} by using the permutation relations in Eq.~(\ref{permutation}).

\begin{algorithm}[!h]
\floatname{algorithm}{Subroutine}
\caption{ Exponentiation of matrix multiplication} \label{algoproduct}
\begin{flushleft}
\textbf{Input:} 
A set of unitary operators according to Input Assumption \ref{oracleUnitaries} for $A_j$ with $j=1,2$, a desired error parameter $\epsilon_2$, desired simulation time $t$, and the unitary operator $U_1$
\\
\textbf{Output:} Operator 
$\mathcal U_{\rm mult}(t)$
which satisfies 
\begin{equation}
 \left \Vert \mathcal U_{\rm mult}(t) -e^{iX_3(A_1A_2)t}\right \Vert \le \epsilon_2.
 \end{equation}
\textbf{Procedure:}  Construct $e^{\pm iX_1(A_1)t/n}$, $e^{\pm iX_2(A_2)t/n}$ via permutation, see Eq.~(\ref{permutation}), and apply them and $U_1$ as follows
\begin{equation}
\mathcal U_{\rm mult}(t)= U_1[\tilde{l}(iX_1(A_1)t/n,iX_2(A_2)t/n)]^{n'}U_1^\dagger,
\end{equation}
where $n'=n^2/2t$ and $n$ is chosen such that $n'=O(t^2/\epsilon_2)$.
\end{flushleft}
\end{algorithm}

%----------------------------------------------------------------------------------
%%  Kronecker sum and tensor product 
%----------------------------------------------------------------------------------

\subsection{Kronecker sum and tensor product }

Building on the above results, we can also find an approximation of the exponentiation of the Kronecker sum and tensor product of matrices $A_1$ and $A_2$. 
Simulating Kronecker sums on quantum computers is the topic of the initial works on quantum simulation \cite{Lloyd1996}, where the Hamiltonian is a sum of local Hamiltonians. For completeness, we present the simulation of Kronecker sums here in the context of the embedding matrices.
The Kronecker sum of these two matrices denoted by $\oplus$ is defined as
\begin{align}
A_1\oplus A_2:=A_1\otimes I +I\otimes A_2,
\end{align}
where $I$ is the identity matrix.  Since $A_1\otimes I $  and $I\otimes A_2$ commute, one can obtain the exponentiation of the Kronecker sum by 
\begin{align}
e^{A_1\oplus A_2}= (e^{A_1}\otimes I) (I\otimes e^{ A_2}). 
\end{align}
Now we consider the exponentiation of $A_1\oplus  A_2$  with the embedding matrices by using subroutine \ref{algoadd}. The exponentiation of embedded $A_1\oplus A_2$ is 
\begin{align}
 e^{iX_3(A_1\oplus A_2)t}=e^{i(X_3(A_1\otimes I) +X_3(I\otimes A_2))t}. \label{kronecker sum}
\end{align}
Using the definition of $X_3(A)$, we have
\begin{eqnarray}
X_3(A_1\otimes I)&=&
R_3\otimes A_1\otimes I+R_3^\dagger \otimes A_1^\dagger\otimes I,\nonumber\\
X_3(I\otimes A_2)&=& R_3\otimes I\otimes A_2+R_3^\dagger \otimes I\otimes A_2^\dagger. \nonumber 
\label{k sum}  
\end{eqnarray}
Considering the three registers,  note that $X_3(A_1)$ acts on the first and second register and $X_3(A_2)$ acts on first and third register.   
Thus, we can approximate $e^{ {iX_3(A_1\oplus A_2)t} }$
using subroutine \ref{algoadd} with the inputs $e^{i{X_3(A_1)\tau}}$ and $e^{iX_3(A_2)\tau}$ on the appropriate registers.

We can also derive the exponentiation of the tensor product of matrices $A_1$ and $A_2$, using the permutation relations in Eq.~(\ref{permutation}) and subroutine \ref{algoproduct}. Since
\begin{align}
{A_1\otimes A_2}=(A_1\otimes I )(I\otimes A_2),
\end{align}
the Hamiltonian simulation of $A_1\otimes A_2$ can be performed with the embedding $e^{iX_3(A_1\otimes A_2)t}$ using subroutine \ref{algoproduct} with the inputs $e^{\pm iX_3(A_1\otimes I)\tau}$ and $e^{\pm iX_3(I \otimes A_2)\tau}$, noting that $X_1(A)$ and $X_2(A)$ can be obtained from $X_3(A)$ via the permutations defined in Eq.~(\ref{permutation}).

%----------------------------------------------------------------------------------
%%  Hadamard product
%----------------------------------------------------------------------------------

\subsection{Hadamard product}

We denote the Hadamard product of matrices $A_1$ and $A_2$  as $ A_1\circ A_2$, which is a matrix given by
\begin{align}
(A_1\circ A_2)_{ij}= ({A_1})_{ij} (A_2)_{ij}.
\end{align}
We now discuss a method to obtain such a Hadamard product in the QMAT framework. Define the following (non-Hermitian) matrix 
\begin{align}
{S}=\textstyle\sum\nolimits_{i=0}^{N-1} \ket i \bra i \otimes\ket  {\bar 0} \bra i,
\end{align}
where $ \ket {\bar 0}:=\ket{0\dots 0}$. Since the matrix $S$ is sparse, there exists an efficient quantum algorithm to simulate the embedded sparse matrix $X_3(S)$ \cite{low2017optimal}.
We can obtain the Hadamard product of $A_1$ and $A_2$ from the tensor product in the following way
\begin{align}
{S}(A_1\otimes A_2){S}^\dagger
=(A_1 \circ A_2)\otimes \ket {\bar 0} \bra {\bar 0}.\label{hadamard product}
\end{align}
The size of the resulting matrix is extended by the operator $\ket{\bar 0}\bra {\bar 0}$. In the embedding matrix form, we note the relation
\begin{eqnarray}
&e^{iX_3({S}(A_1\otimes A_2){S}^\dagger)t} 
=e^{iX_3(A_1 \circ  A_2)\otimes\ket{\bar 0}\bra{\bar 0} t}\nonumber\\
&\quad \quad= e^{iX_3(A_1 \circ  A_2)t}\otimes\ket{\bar 0}\bra{\bar 0}+ I\otimes(I-\ket{\bar 0}\bra{\bar 0}),
\end{eqnarray}
where the last equality follows from the series expansion of the exponential. 
Thus, we approximate  $e^{iX_3(A_1 \circ A_2)t}$  by using $e^{iX_3({S}(A_1\otimes A_2){S}^\dagger)t}$ with an ancillary register in state $ \ket {\bar 0}$. 
The operator $e^{iX_3({S}(A_1\otimes A_2){S}^\dagger)t}$ can be constructed by  
combining the methods for the matrix tensor product and the multiplication subroutine \ref{algoproduct} with access to $e^{\pm iX_3({S})\tau}$, $e^{\pm iX_3({S}^\dagger)\tau}$, $e^{\pm iX_3(A_1)\tau}$ and $e^{\pm iX_3(A_2)\tau}$. 

%----------------------------------------------------------------------------------
%%  sub-functions of arbitrary real operator functions
%----------------------------------------------------------------------------------

\subsection{Real operator functions}

We define an arbitrary real operator function $h(A)$ of a Hermitian matrix $A$ as
 \begin{equation}
 h(A)=\sum_{\lambda}h(\lambda)\ket{\mu_{\lambda}}\bra{\mu_{\lambda}},
 \end{equation}
 where $\lambda$ and $\ket{\mu_{\lambda}}$ are the eigenvalues and the corresponding eigenvectors respectively.
Such a function can be applied in the QMAT framework using results presented in Ref.~\cite{low2017optimal,Gilyen2018}, where a procedure was given to construct a quantum circuit that performs $ W'=\sum_{\lambda}e^{ih(\lambda)}\ket{u_{\lambda}}\bra{u_{\lambda}}$ for a real odd and periodic function $h(\lambda):(-\pi,\pi]\rightarrow (-\pi,\pi]$, given a controlled unitary $W=\ket{0}\bra{0}\otimes I + \ket{1}\bra{1}\otimes \sum_{\lambda}e^{i\lambda}\ket{u_{\lambda}}\bra{u_{\lambda}}$. One can apply the formalism to even functions $h(\lambda)$ by simply dividing the even function into two odd functions  $h(\lambda)/\lambda$ and ${\lambda}$, then multiplying them with using the functionality of  subroutine \ref{algoproduct}. More generally, if $h(\lambda)$ is neither even nor odd, we first divide it into an even function $ \frac{1}{2}(h(\lambda)+h(-\lambda))$ and an odd function $\frac{1}{2}(h(\lambda)-h(-\lambda))$, then recombine them using the functionality of subroutine \ref{algoadd}. 

Now we consider an odd  function $h(A)$ of a Hermitian matrix $A$ in the embedding formula. According to the following equation
\begin{eqnarray}
\left[\begin{array}{ccc}
0 & 0 & A\\
0 & 0 & 0\\
A & 0 & 0 
\end{array} \right]
\left[\begin{array}{c}
\ket{u_{\lambda} } \\
0 \\
\pm \ket{u_{\lambda} } 
\end{array} \right]
=\pm \lambda
\left[\begin{array}{c}
\ket{u_{\lambda} } \\
0 \\
\pm \ket{u_{\lambda} } 
\end{array} \right], 
\end{eqnarray}
we see that the eigenvalues of $X_3(h(A)) $ are $\pm h(\lambda)$ whereas the eigenvalues of $h(X_3(A))$ are $h(\pm \lambda)$. Thus we have ${X_3(h(A))}={h(X_3(A))}$ since $ \pm h(\lambda)=h(\pm \lambda)$ for an odd function. 

As such we have shown that given an input of the form $e^{iX_3(A)\tau}$, it is possible to obtain $e^{iX_3(h(A))t}$ for general real operator functions. The outcome of this subroutine is in the desired matrix exponent form, and hence can be concatenated with all other subroutines discussed in this paper.

To summarize, these subroutines allow performing a mix of operations of matrices in a concatenated fashion. Given a set of  unitaries generated by $\{X_3(A_j)\}$, $j=1, \dots, J$, and a large class of functions $ f(\{A_j\})$ of these matrices, 
we are able to construct the unitary operator $e^{iX_3(f(\{A_j\}))t}$. We note that each procedure produces an output satisfying the Input Assumption \ref{oracleUnitaries}, which allows for concatenation of the procedures. The value of $\bra{x} f\left(\{A_j\}\right)\ket{y}$ can be estimated efficiently, as was shown in Sections \ref{secOutput}. Other scalar quantities can also be estimated, as will be shown in the next section.

%----------------------------------------------------------------------------------
%% trace 
%----------------------------------------------------------------------------------

\subsection{Estimating norms, traces and determinants}
Let $A=f(\{A_j\})$, we now show a method to approximate the Schatten $p$-norm of $A$ using the QMAT embedding matrices. The Schatten $p$-norm of matrix $A$ is defined as $\Vert A\Vert_p=(\sum_{k=1}^{N} \sigma_k^p)^{1/p}$ for  $\sigma_1 \geq \sigma_2...\geq \sigma_N \geq 0$ the singular values of $A$. Since the eigenvalues of $X_3(A)$ are $\{\pm \sigma_k\}$, we can estimate the Schatten $p$-norm by sampling the absolute value of the p-th power of the eigenvalues of $X_3(A)$ then calculating the $p$-th root of the result.  It can be achieved by performing $e^{iX_3(A)}$ on a maximally mixed state then using phase estimation and measurement to get the eigenvalues. A discussion on the Schatten $p$-norm via the DQC1 protocol can be found in \cite{Cade2017}.

Similarly, the trace of a Hermitian matrix can be estimated by sampling the eigenvalues. Notice that, in the embedding formula, the trace of $X_3(A)$ is zero, thus we cannot estimate the trace of $A$ from the eigenvalues of $X_3(A)$ directly. If $A$ is non-Hermitian, we construct the Hermitian matrices $A+A^\dagger$ and $i(A-A^\dagger)$. Then the real part of $\Tr(A)$ is equal to $\frac{1}{2}\Tr(A+A^\dagger)$ and the imaginary part is equal to $\frac{i}{2}\Tr(A-A^\dagger)$.
Now we show the method to estimate $\Tr(A)$ where $A$ is Hermitian. 
The eigenvalues of $A$ are $\{\lambda_i\}$, such that the eigenvalues of $X_3(A)$ are $\{\pm \lambda_i\}$. Construct a matrix $A'=A+cI$, with $c$ such that all eigenvalues of $A$ are shifted to positive values. 
The sampling method proceeds as follows: apply $e^{iX_3(A')}$ on a maximally mixed state, then perform phase estimation and measurement which yield one of the eigenvalues of $X_3(A')$, say $\pm \lambda'_i = \pm (\lambda_i + c)$. Using the absolute value of the measurement outcome, the corresponding eigenvalue of $A$ can be extracted by $\lambda_i=\vert \lambda'_i\vert -c $, which is then used to evaluate the desired trace. This procedure works, because by operating on the maximally mixed state we are effectively choosing an eigenvector uniformly at random, and hence sampling the eigenvalues uniformly at random. An alternate approach to trace estimation would be to sample diagonal elements of the matrix in a basis chosen uniformly at random from a maximal unbiased set. In the classical problem of stochastic trace estimation, such a procedure has been shown to yield the same expectation value with greatly reduced variance when compared to sampling in any fixed basis~\cite{fitzsimons2016improved}.

 If $A$ is a Hermitian positive definite matrix, one can estimate the determinant $\det[A]$ as well. This can be achieved by leveraging the equality $\log(\det[A])={\Tr(\log(A))}$, and hence $\det[A] = \exp(\Tr(\log(A)))$. Hence by combining a trace estimation procedure with a procedure to generate evolution under $\log(A)$, and post processing the result, one obtains a procedure to estimate $\det[A]$~\cite{zhao2018quantum}. The error analysis of the estimation methods is given in Appendix \ref{error analysis}. It shows that the Schatten-$p$ norms, the trace and determinant of matrix $A$ can be estimated with small relative error with high success probability.

\subsection{Example}

\begin{figure}[htbp] 
\centering
\includegraphics[width=0.8\columnwidth]{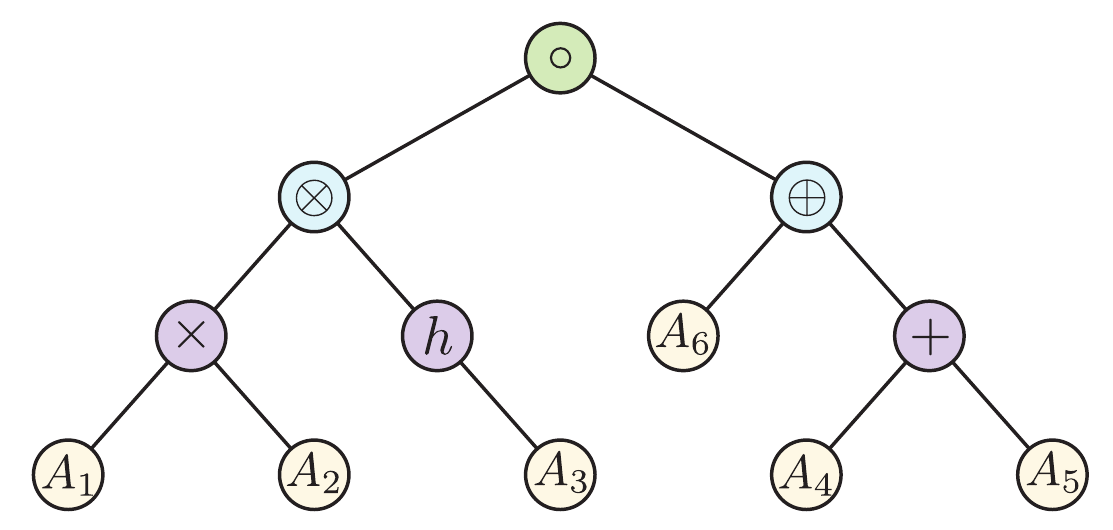} 
\caption{Example of an abstract syntax tree for the function $f(\{A_j\})=((A_1A_2)\otimes h(A_3))\circ (A_6 \oplus (A_4+A_5))$. }\label{figure1}
\end{figure}

Given a matrix function $f(\{A_j\})$, we express the function $f(\cdot)$ as an abstract syntax tree, composed of basic single-matrix and two-matrix operations. Consider an example in which we have six matrices, $A_j$ for $j=1,\dots,6$, and a matrix function $f(\{A_j\})=((A_1A_2)\otimes h(A_3))\circ (A_6 \oplus (A_4+A_5))$. Figure \ref{figure1} shows the corresponding tree. The unitary  $e^{iX_3(f(\{A_j\}))}$ is generated by concatenating the different subroutines using the embedding matrices.

%---------------------------------------------------------------------
%Conclusion 
%---------------------------------------------------------------------

\section{Discussion and conclusion}

The Quantum Matrix Algebra Toolbox that we have presented here gives a consistent and composable set of procedures for manipulating matrices, represented as unitary evolution of a quantum system, and for estimating scalar valued functions of these matrices. This toolbox allows for the estimation of scalar properties of matrix functions corresponding to arbitrary abstract syntax trees containing real eigenvalue functions of a single matrix as well as the most common two-matrix products and sums. Matrix embeddings allow us to simulate complex unitaries with sequences of simple unitaries, in the spirit of the usual Trotter-based Hamiltonian simulation techniques. For a set of matrices $\{A_j\}$, the methods can obtain $\bra{x} f(\{A_j\}) \ket{y}$  and the trace of $f(\{A_j\})$ with bounded errors, and also individual elements of $f(\{A_j\})$ if we choose $\bra{x}$ and $\ket{y}$ as the row and column basis vectors.  As the underlying sparse Hamiltonian techniques are logarithmic in the dimension of the matrices, we may achieve the estimation of such quantities much faster than classical algorithms which often scale polynomial in the dimension.

In our approach, we have built up the full set of matrix operations by using Trotter formulae to approximate evolution under the sum or commutator of matrices. Alternative approaches exist for calculating matrix functions with exponentially improved error scaling. In Ref.~\cite{Gilyen2018} it was shown that one can compute addition and multiplication of two matrices if the block-encodings of these matrices are given. The techniques in the present work can potentially be combined with the block-encoding framework to construct alternative compiling schemes with improved error scaling and hence improved efficiency.

In conclusion, the techniques we present here represent a systematic method for the compilation of user-specified instruction sets of desired matrix operations into gate sequences for actual quantum computers. We believe that these techniques will prove useful in constructing further quantum algorithms based on fast matrix algebra.

\acknowledgments

This material is based on research supported by the Singapore National Research Foundation under grant NRF-NRFF2013-01 and NRF2017-NRF-ANR004, and by the US Airforce Office of Scientific Research under grant FA2386-18-1-4003. The authors also acknowledge support from Singapore's Ministry of Education.

\appendix 

\section{Commutator simulation}
\label{appendixHS}

Given access to $e^{ H_1t/n}$ and $e^{H_2t/n}$  for two  matrices $H_1, H_2 \in \mathbb{C}^{N\times N}$ for small $t$, we can approximate $e^{[H_1,H_2]t}$ with bounded error by using the second order of the Baker-Campbell-Hausdorff formula \cite{blanes2004convergence} as follows
\begin{eqnarray}
 e^{[H_1,H_2]t+O({ {{t^2}/{n^2}}})}
= (e^{H_1{t/n}}e^{H_2 {t/n}}e^{-H_1 {t/n}}e^{-H_2 {t/n}})^{n^2/t}.\nonumber \\\label{commute}
\end{eqnarray} 
In order to reduce the error, we rearrange the order of  every term in the right side of  Eq.~(\ref{commute}). 
Now we define   
 \begin{eqnarray}
l(H_1t/n,H_2t/n):=e^{H_1t/n}e^{H_2t/n}e^{-H_1t/n}e^{-H_2 t/n},
 \end{eqnarray} 
 and we have
\begin{eqnarray}
  l(-H_1t/n,-H_2t/n)&=&e^{-H_1t/n}e^{-H_2t/n}e^{H_1t/n}e^{H_2t/n}.\nonumber \\
\end{eqnarray}
 Then, combining the above two equations, we obtain
\begin{eqnarray}
\tilde{l}(H_1 t/n, H_2 t/n)&:=&l(H_1 t/n, H_2 t/n)l(-H_1 t/n, -H_2 t/n)\nonumber\\
&=& e^{2[H_1,H_2]t^2/n^2+O({(t/n)^{4}})}.
\end{eqnarray}
Hence, the term  $e^{[H_1, H_2]t}$ can be approximated with bounded error as
  \begin{equation}
  \tilde{l}(H_1t/n,H_2t/n)^{n^2/2t}=e^{[H_1,H_2]t+O(t^{3}/n^2)}.\label{equmultiplication}
  \end{equation}
Let $n'=n^2/2t$, 
we see that $e^{[H_1,H_2]t}$ can be approximated to a constant $\epsilon_2$-error by using $n'=O({t^{2}/\epsilon_2})$ copies of $e^{H_1t/n}$ and $e^{H_2t/n}$.

\section{Quantum inner product estimation}
\label{appendixQIP}

We now review a method for obtaining $\braket{x}{y}$ for given vectors $\ket{x},\ket{y}\in\mathbb{C}^N$. Assume we are give a state 
\begin{eqnarray}
\ket{\varphi}=\frac{1}{\sqrt{2}}(\ket{0}\ket{x}+\ket{1}\ket{y})\label{inner product}.
\end{eqnarray}
Applying a Hadamard operator on the first qubit, this yields
\begin{eqnarray}
\ket{\varphi}=\frac{1}{{2}}(\ket{0}(\ket{x}+\ket{y})+\ket{1}(\ket{x}-\ket{y})).
\end{eqnarray}
Measuring the first qubit in computational basis, the probability to obtain $\ket{0}$ is given by
\begin{eqnarray}
p=\frac{1}{2}(1+\rm Re(\braket{x}{y})).
\end{eqnarray}
By repeating this procedure a constant number of times, we can estimate the real part of $\braket{x}{y}$ to fixed precision. Likewise, the imaginary part of $\braket{x}{y}$ can be obtained by applying a phase rotation on the first register in Eq.~(\ref{inner product}), to produce
\begin{eqnarray}
\ket{\varphi}=\frac{1}{\sqrt{2}}(\ket{0}\ket{x}-i\ket{1}\ket{y}),
\end{eqnarray} 
followed by applying a Hadamard matrix and measuring the first qubit in Pauli $Z$ basis, the probability of obtaining $\ket{0}$ is given by 
\begin{eqnarray}
p=\frac{1}{2}(1+\rm Im(\braket{x}{y})).
\end{eqnarray}
This can be repeated a constant number of times to get an estimate of the imaginary part of $\braket{x}{y}$ to fixed precision. In this way, we can approximate the inner product of vector $\bold{x}$ and $\bold{y}$.
 
As the variance of binomial distribution specified by the probability $p$ and the number of trials $m$ %$B(M,p)$ 
is given by $mp(1-p)$, the variance of the estimate for $p$ is equal to $p(1-p)/m$. The variance of the real part is $4$ times the variance of the estimate for $p$. Thus the error of the real part of the inner product is then given by 
\begin{equation}
\left (\frac{1}{m}(1-\rm Re(\braket{x}{y})^2) \right)^{\frac{1}{2}},
\end{equation}
 which is then bounded by $ \frac{1}{\sqrt{m}}$, and similarly for the imaginary part.

\section{Error analysis of the Schatten-p norm estimation}
\label{error analysis}

We would like to estimate the Schatten-p norm of a matrix $A$ by uniform sampling its singular values. 
Assume a relative error of the singular values $\sigma_i$ from quantum phase estimation be $\epsilon_\sigma>0$.
Concretely, we are interested in estimating $ \Vert A \Vert ^p_p$ to fixed accuracy, independent of the dimension of $A$, with high probability.

Formally, we have random variables $\tilde \sigma_i$ with expectation value $E[\tilde \sigma_i] = \sigma_i$ and variance $\text{Var}[\tilde \sigma_i] = \epsilon_\sigma^2 \sigma_i^2$.
In addition, we sample the index $i$ uniformly, thus we define a random variable $\tilde X$ to be the random variable that has as outcomes each $\tilde \sigma_i^p$ with probability $1/N$. Let $K_p$ the Lipshitz constant for the function $f_p(x) = x^p$ in the interval $[0,\sigma_{\max}]$. Note that
\begin{equation}
\vert \sigma_i^p - \tilde \sigma_i^p \vert \leq K_p \vert \sigma_i - \tilde \sigma_i \vert,
\end{equation} 
and also the special case $\tilde \sigma_i^{2p} \leq K_p^2  \tilde \sigma_i^2 $.
Let $\tilde Y$ be the random variable $\frac{N}{T}\sum_{\rm j = 1}^T \tilde X_j$, where $\tilde X_j$ denote independent instances of the random variable $\tilde X$.
This random variable ($\tilde Y$) has the expectation value, 
\begin{eqnarray}
E[\tilde Y] &=&\frac{N}{T} E\left[\sum_{j=1}^T \tilde X_j\right]\nonumber\\
&=&N E[\tilde X] = N \sum_{j=1}^N \frac{1}{N} E\left[\tilde \sigma_j^p\right] \nonumber\\
&=&\Vert A\Vert_p^{p}+ \sum_{j=1}^N E[\tilde \sigma_j^p]-\Vert A\Vert_p^{p} \nonumber \\ 
&\leq& \Vert A\Vert_p^{p}+ \sum_{j=1}^N E[\vert \tilde \sigma_j^p - \sigma^p_j\vert ]  \nonumber\\
&\leq &\Vert A\Vert_p^{p}+  K_p \sum_{j=1}^N  E[\vert \tilde \sigma_j -\sigma_j\vert ] \nonumber\\
&\leq &\Vert A\Vert_p^{p}+ K_p \sum_{j=1}^N  \sqrt{E[\vert \tilde \sigma_j -\sigma_j\vert^2 ]} \nonumber\\
&= &\Vert A\Vert_p^{p}+ K_p \sum_{j=1}^N  \sqrt{\text{Var}[\tilde \sigma_j]} \nonumber\\
&\leq &\Vert A\Vert_p^{p}+ K_p \sum_{j=1}^N  \epsilon_\sigma \sigma_i \nonumber\\
&=&\Vert A\Vert_p^{p}+K_p \epsilon_\sigma\Vert A\Vert_1.
\end{eqnarray}
Here, we have used Jensen's inequality twice and Lipshitz continuity.
For the quadratic term, we have, 
\begin{eqnarray}
E[\tilde Y^2]&=& \frac{N^2}{T^2} E\left[ \sum_{j=1}^T \tilde X_j^2 \right ]  \nonumber \\ 
&=& \frac{N^2}{T} E\left[ \tilde X^2 \right ] \nonumber\\
&=& \frac{N}{T} \sum_{j=1}^N E\left[ \tilde \sigma_j^{2p} \right ] \nonumber \\
&\leq &  \frac{N K_p^2}{T} \sum_{j=1}^N  E\left[ \tilde \sigma_j^{2} \right ] \nonumber\\
&=&\frac{N K_p^2 (1+\epsilon_\sigma^2)}{T} \sum_{j=1}^N \sigma_j^2 \nonumber \\
&=& \frac{N K_p^2 (1+\epsilon_\sigma^2)}{T} \Vert A \Vert_2^2 
\end{eqnarray}

According to Chebyshev's inequality, we obtain
\begin{eqnarray}
P\left(\vert \tilde Y- E[\tilde Y]\vert \geq \varepsilon \Vert A\Vert_{p}^p \right) &\leq& \frac{\text{Var}(\tilde Y)}{\varepsilon^2 \Vert A\Vert_{p}^{2p}}
 \leq  \frac{E(\tilde Y^2)}{\varepsilon^2 \Vert A\Vert_{p}^{2p}} \nonumber \\
 &\leq& \frac{N}{T \varepsilon^2} \frac{K_p^2 (1+\epsilon_\sigma^2)}{\Vert A\Vert_{p}^{2p}} \Vert A \Vert_2^2. \nonumber
\end{eqnarray}
This failure probability is employed to bound the final error. 
Note that 
\begin{equation}
\vert \tilde Y - \Vert A\Vert_p^{p} \vert \leq \vert \tilde Y - E[\tilde Y]  \vert + K_p \epsilon_\sigma\Vert A\Vert_1.
\end{equation}
Thus, with probability $1-P_\text{fail}$ the error is bounded as
\begin{equation}
\vert \tilde Y - \Vert A\Vert_p^{p} \vert \leq \varepsilon \Vert A\Vert_{p}^p + K_p \epsilon_\sigma\Vert A\Vert_1,
\end{equation}
where
\begin{equation}
P_{\rm fail} \leq \frac{N}{T \varepsilon^2} \frac{K_p^2 (1+\epsilon_\sigma^2)}{\Vert A\Vert_{p}^{2p}} \Vert A \Vert_2^2.
\end{equation}
Assume that $\sigma_i=\Theta(1)$ for all $i$. Then the above expression can be simplified as
\begin{eqnarray}
P_\text{fail} \leq  \frac{C K_p^2 (1+\epsilon_\sigma^2)  }{T \varepsilon^2}.
\end{eqnarray} 
where $C$ is a constant. Let $\frac{C K_p^2 (1+\epsilon_\sigma^2)  }{T \varepsilon^2}=a$, then
the number of samples required is $T = \left \lceil \frac{C K_p^2 (1+\epsilon_\sigma^2)  }{\varepsilon^2 a } \right \rceil$. We can take $a=0.01$, for example, to achieve a $99\%$ confidence.
In summary, the outcome of the estimation procedure is bounded by
\begin{eqnarray}
\Vert A\Vert_{p} \left(1\pm \varepsilon \pm K_p \epsilon_\sigma\frac{\Vert A\Vert_1}{\Vert A\Vert_{p}^{p}} \right)^{\frac{1}{p}},\label{outcome p-norm} 
\end{eqnarray} 
with high probability. Since $x^{1/p}$ is a concave function, and since we have already assumed that the singular values satisfy $\sigma_i=\Theta(1)$, this implies that the relative error in the estimation of $\Vert A\Vert_{p}$ is bounded by 
\begin{eqnarray}
\varepsilon_R &\leq & |\varepsilon|^{\frac{1}{p}} + \left(K_p |\epsilon_\sigma| \frac{\Vert A\Vert_1}{\Vert A\Vert_{p}^{p}}\right)^{\frac{1}{p}}\\
&=& |\varepsilon|^{\frac{1}{p}} + \kappa_p |\epsilon_\sigma|^\frac{1}{p},\label{relative error} 
\end{eqnarray} 
where $\kappa_p$ is some constant that depends only on $p$ and the allowed (constant) range of the singular values of $A$. Thus the relative error does not depend on the dimensions of $A$, provided that the assumption on the relative size of the non-zero singular values of $A$ holds. 

The error analysis of trace estimation is similar to the Schatten $1$-norm. Assume the relative error from phase estimation of the eigenvalues be $\varepsilon_\lambda$. The sampling outcome is then bounded by $\Tr(A)(1\pm \varepsilon \pm  \epsilon_\lambda)$ with a number of samples $T= \frac{C(1+\epsilon_\lambda^2)}{\varepsilon^2 a}$, which is obtained from Eq. (\ref{outcome p-norm}) taking $K_p=1$ and $p=1$. 

Finally, for the determinant, the error analysis is as follows. Let $\epsilon_{\log(\lambda)}$ be the relative error of the eigenvalues of $\log(A)$ from the phase estimation step. Since $\det(A)=e^{\Tr(\log(A))}$, the determinant of a matrix $A$ can be estimated with a number of samples $T=\frac{C(1+\epsilon_{\log(\lambda)}^2)}{\varepsilon^2 a}$. The error is then given by
\begin{eqnarray}
\vert e^{\Tr(\log(A))}-e^{\widetilde{\Tr(\log(A))}}\vert 
&\leq & K_e \vert {\Tr(\log(A))}-{\widetilde{\Tr(\log(A))}}\vert\nonumber\\
& = & K_e \vert \Tr(\log(A))(\epsilon_{log(\lambda)}+\epsilon)\vert, \nonumber
\end{eqnarray}
with high probability, where $K_e$ is the Lipshitz constant for the function $f(x)=e^{x}$. The Lipshitz constant is $1$ if $x \leq 0$, which corresponds to the case when the spectral norm is $\Vert A\Vert \leq 1$.

\bibliographystyle{apsrev}
\bibliography{QBLAS}

\end{document}